\documentclass{LMCS}

\usepackage{amssymb,latexsym,amsmath,wrapfig,comment,enumerate,hyperref}

\newcommand\Meas[1]{\MM(#1)}
\newcommand\norm[1]{{\parallel}#1{\parallel}}

\newcommand\argmax{\mathrm{argmax}}
\newcommand\Exp[1]{\EE\kern1pt{#1}}

\newcommand\one{\mathbf{1}}
\newcommand\xpose{^{\mathrm{T}}}

\newcommand\head{\ensuremath{\mathsf{head}}}
\newcommand\tail{\ensuremath{\mathsf{tail}}}
\newcommand\Obs{\ensuremath{\mathsf{obs}}}
\newcommand\Cont{\ensuremath{\mathsf{cont}}}

\newcommand\sP{P - \frac 1n\one\one\xpose}
\newcommand\APG{\ensuremath{\mathrm{APG}}}
\newcommand\HF{\ensuremath{\mathrm{HF}}}
\newcommand\HC{\ensuremath{\mathrm{HC}}}

\newcommand\Hphi{\ensuremath{\mathrm{H}\phi}}
\newcommand\cl{\ensuremath{\mathrm{cl}}}
\newcommand\powerset[1]{2^{#1}}

\newcommand\eps{\varepsilon}
\newcommand\imp{\rightarrow}
\newcommand\Imp{\Rightarrow}
\newcommand\Iff{\Leftrightarrow}
\newcommand\subs{\,\subseteq\,}
\newcommand\set[2]{\{#1 \mid #2\}}
\newcommand\eqdef{\stackrel{\rm def}{=}}
\newcommand\Iffdef{\stackrel{\rm def}{\Longleftrightarrow}}
\newcommand\reals{{\mathbb R}}
\newcommand\len[1]{|#1|}
\newcommand\MM{{\mathbb M}}

\newlength\reasonwidth
\newcommand\reasoning[1]{\def\longest{#1}\settowidth{\reasonwidth}{$\displaystyle\longest$}\addtolength{\reasonwidth}{5mm}} %dck 2/12/98
\newcommand\reason[2]{\makebox[\reasonwidth][l]{$\displaystyle{#1}$}\mbox{#2}}
\renewcommand\star{^{\textstyle *}}
\newcommand\union{\mathrel{\cup}}
\newcommand\cir{\circle*{4}}
\renewcommand\phi{\varphi}
\renewcommand\emptyset\varnothing
\theoremstyle{plain}

\def\doi{5 (3:10) 2009}
\lmcsheading%
{\doi}
{1--19}
{}
{}
{Dec.~20, 2007}
{Sep.~16, 2009}
{}   

\begin{document}

\title[Applications of Metric Coinduction]{Applications of Metric Coinduction\rsuper*}

\author[D.~Kozen]{Dexter Kozen\rsuper a}	%required
\address{{\lsuper a}Computer Science Department, Cornell University, Ithaca, NY 14853-7501, USA}	%required
\email{kozen@cs.cornell.edu}  %optional
%\thanks{Partially supported by NSF grant CCF-0635028.}	%optional

\author[N.~Ruozzi]{Nicholas Ruozzi\rsuper b}	%optional
\address{{\lsuper b}Computer Science Department, Yale University, New Haven, CT 06520-8285, USA}	%optional
\email{Nicholas.Ruozzi@yale.edu}  %optional
%\thanks{thanks 2, optional.}	%optional

%% etc.

%% required for running head on odd and even pages, use suitable
%% abbreviations in case of long titles and many authors:

%% mandatory lists of keywords and classifications:
\titlecomment{{\lsuper*}A preliminary version of this paper appeared as \cite{KR07a}.}
%%%%%%%%%%%%%%%%%%%%%%%%%%%%%%%%%%%%%%%%%%%%%%%%%%%%%%%%%%%%%%%%%%%%%%%%%%%

%% the abstract has to PRECEED the command \maketitle:
%% be sure not to issue the \maketitle command twice!

\begin{abstract}
\noindent Metric coinduction is a form of coinduction that can be used to establish properties of objects constructed as a limit of finite approximations.  One can prove a coinduction step showing that some property is preserved by one step of the approximation process, then automatically infer by the coinduction principle that the property holds of the limit object.  This can often be used to avoid complicated analytic arguments involving limits and convergence, replacing them with simpler algebraic arguments.  This paper examines the application of this principle in a variety of areas, including infinite streams, Markov chains, Markov decision processes, and non-well-founded sets.  These results point to the usefulness of coinduction as a general proof technique.
\end{abstract}

\maketitle

\section{Introduction}
\label{sec:intro}

Mathematical induction is firmly entrenched as a fundamental and ubiquitous proof principle for proving  properties of inductively defined objects.  Mathematics and computer science abound with such objects, and mathematical induction is certainly one of the most important tools, if not the most important, at our disposal.

Perhaps less well entrenched is the notion of coinduction.  Despite recent interest, coinduction is still not fully established in our collective mathematical consciousness.  A contributing factor is that coinduction is often presented in a relatively restricted form.  Coinduction is often considered synonymous with bisimulation and is used to establish equality or other relations on infinite data objects such as streams \cite{Rutten00} or recursive types \cite{fiore93coinduction}.

In reality, coinduction is far more general.  For example, it has been recently been observed \cite{K07b} that coinductive reasoning can be used to avoid complicated $\eps$-$\delta$ arguments involving the limiting behavior of a stochastic process, replacing them with simpler algebraic arguments that establish a \emph{coinduction hypothesis} as an invariant of the process, then automatically deriving the property in the limit by application of a coinduction principle.  The notion of bisimulation is a special case of this: establishing that a certain relation is a bisimulation is tantamount to showing that a certain coinduction hypothesis is an invariant of some process.

Coinduction, as a proof principle, can handle properties other than equality and inequality and extends to other domains.  The goal of this paper is to explore some of these applications.  We focus on four areas: infinite streams, Markov chains, Markov decision processes, and non-well-founded sets.  In Section \ref{sec:coind}, we present the metric coinduction principle.  In Section \ref{sec:streams}, we illustrate the use of the principle in the context of infinite streams as an alternative to traditional methods involving bisimulation.  In Sections \ref{sec:mc} and \ref{sec:mdp}, we rederive some basic results of the theories of Markov chains and Markov decision processes, showing how metric coinduction can simplify arguments.  Finally, in Section \ref{sec:nwf}, we use metric coinduction to derive a new characterization of the hereditarily finite non-well-founded sets.

\section{Coinduction in Complete Metric Spaces}
\label{sec:coind}

\subsection{Contractive Maps and Fixpoints}
\label{sec:contractive}

Let $(V,d)$ be a complete metric space.  A function $H:V\imp V$ is \emph{contractive} if there exists $0\leq c < 1$ such that for all $u,v\in V$, $d(H(u),H(v)) \leq c\cdot d(u,v)$.  The value $c$ is called the \emph{constant of contraction}.  A continuous function $H$ is said to be \emph{eventually contractive} if $H^n$ is contractive for some $n\geq 1$.  Contractive maps are uniformly continuous, and by the Banach fixpoint theorem, any such map has a unique fixpoint in $V$.

The fixpoint of a contractive map $H$ can be constructed explicitly as the limit of a Cauchy sequence $u,H(u),H^2(u),\ldots$ starting at any point $u\in V$.  The sequence is Cauchy; one can show by elementary arguments that
\begin{eqnarray*}
d(H^{n+m}(u),H^n(u)) &\leq& c^n(1-c^m)(1-c)^{-1}\cdot d(H(u),u).
\end{eqnarray*}
Since $V$ is complete, the sequence has a limit $u\star$, which by continuity must be a fixpoint of $H$.  Moreover, $u\star$ is unique: if $H(u)=u$ and $H(v)=v$, then
\begin{eqnarray*}
d(u,v) = d(H(u),H(v)) \leq c\cdot d(u,v) &\Imp& d(u,v)=0,
\end{eqnarray*}
therefore $u=v$.

Eventually contractive maps also have unique fixpoints.  If $H^n$ is contractive, let $u\star$ be the unique fixpoint of $H^n$.  Then $H(u\star)$ is also a fixpoint of $H^n$.  But then $d(u\star,H(u\star)) = d(H^n(u\star),H^{n+1}(u\star)) \leq c\cdot d(u\star,H(u\star))$, which implies that $u\star$ is also a fixpoint of $H$.

\subsection{The Coinduction Rule}
\label{sec:rule}

In the applications we will consider, the coinduction rule takes the following simple form: If $\phi$ is a closed nonempty subset of a complete metric space $V$, and if $H$ is an eventually contractive map on $V$ that preserves $\phi$, then the unique fixpoint $u\star$ of $H$ is in $\phi$.  Expressed as a proof rule, this says for $\phi$ a closed property,
\begin{equation}
\frac{\exists u\ \phi(u) \qquad \forall u\ \phi(u) \Imp \phi(H(u))}{\phi(u\star)}.\label{eqn:coind1}
\end{equation}
In \cite{K07b}, the rule was used in the special form in which $V$ was a Banach space (normed linear space) and $H$ was an eventually contractive linear affine map on $V$.

\subsection{Why Is This Coinduction?}
\label{sec:why}

We have called \eqref{eqn:coind1} a coinduction rule.  To justify this terminology, we must exhibit a category of coalgebras and show that the rule \eqref{eqn:coind1} is equivalent to the assertion that a certain coalgebra is final in the category.  This construction was given in \cite{K07b}, but we repeat it here for completeness.

Say we have a contractive map $H$ on a metric space $V$ and a nonempty closed subset $\phi\subs V$ preserved by $H$.  Define $H(\phi)=\set{H(s)}{s\in\phi}$.  Consider the category $C$ whose objects are the nonempty closed subsets of $V$ and whose arrows are the reverse set inclusions; thus there is a unique arrow $\phi_1\imp\phi_2$ iff $\phi_1\supseteq\phi_2$.  The map $\bar H$ defined by $\bar H(\phi)=\cl(H(\phi))$, where $\cl$ denotes closure in the metric topology, is an endofunctor on $C$, since $\bar H(\phi)$ is a nonempty closed set, and $\phi_1\supseteq\phi_2$ implies $\bar H(\phi_1)\supseteq \bar H(\phi_2)$.  An $\bar H$-coalgebra is then a nonempty closed set $\phi$ such that $\phi\supseteq \bar H(\phi)$; equivalently, such that $\phi\supseteq H(\phi)$.  The final coalgebra is $\{u\star\}$, where $u\star$ is the unique fixpoint of $H$.  The coinduction rule \eqref{eqn:coind1} says that $\phi\supseteq H(\phi)\Imp \phi\supseteq\{u\star\}$, which is equivalent to the statement that $\{u\star\}$ is final in the category of $\bar H$-coalgebras.

\section{Streams}
\label{sec:streams}

\renewcommand\SS{\mathcal{S}_\Sigma}
\newcommand\mergename{\mathsf{merge}}
\newcommand\splitname{\mathsf{split}}
\newcommand\merge[2]{\mergename\,(#1,\,#2)}
\renewcommand\split[1]{\splitname\,(#1)}

Infinite streams have been a very successful source of application of coinductive techniques.  The space $\SS = (\Sigma^\omega,\head,\tail)$ of infinite streams over $\Sigma$ is the final coalgebra in the category of \emph{simple transition systems} over $\Sigma$, whose objects are $(X,\Obs,\Cont)$, where $X$ is a set, $\Obs:X\imp\Sigma$ gives an \emph{observation} at each state, and $\Cont:X\imp X$ gives a \emph{continuation} (next state) for each state.  The unique morphism $(X,\Obs,\Cont) \imp (\Sigma^\omega,\head,\tail)$ maps a state $s\in X$ to the stream $\Obs(s),\Obs(\Cont(s)),\Obs(\Cont^2(s)),\ldots\in\Sigma^\omega$.

We begin by illustrating the use of the metric coinduction principle in this context as an alternative to traditional methods involving bisimulation.  It is well known that $\SS$ forms a complete metric space under the distance function $d(\sigma,\tau) \eqdef 2^{-n}$, where $n$ is the first position at which $\sigma$ and $\tau$ differ.  The metric $d$ satisfies the property
\begin{eqnarray*}
d(x::\sigma,y::\tau) &=&
\begin{cases}
\frac 12 d(\sigma,\tau), & \text{if } x = y\\
1, & \text{if } x \neq y.
\end{cases}
\end{eqnarray*}
One can also form the product space $\SS^2$ with metric
\begin{eqnarray*}
d((\sigma_1,\sigma_2),(\tau_1,\tau_2)) &\eqdef& \max d(\sigma_1,\tau_1),\,d(\sigma_2,\tau_2).
\end{eqnarray*}
Since distances are bounded, the spaces of continuous operators $\SS^2\imp\SS$ and $\SS\imp\SS^2$ are also complete metric spaces under the sup metric
\begin{eqnarray*}
d(E,F) &\eqdef& \sup_x d(E(x),F(x)).
\end{eqnarray*}

Consider the operators $\mergename:\SS^2\imp\SS$ and $\splitname:\SS\imp\SS^2$ defined informally by 
\begin{eqnarray*}
\merge{a_0a_1a_2\cdots}{b_0b_1b_2\cdots} &=& a_0b_0a_1b_1a_2b_2\cdots\\
\split{a_0a_1a_2\cdots} &=& (a_0a_2a_4\cdots,\,a_1a_3a_5\cdots).
\end{eqnarray*}
Thus $\mergename$ forms a single stream from two streams by taking elements alternately, and $\splitname$ separates a single stream into two streams consisting of the even and odd elements, respectively.

Formally, one would define $\mergename$ and $\splitname$ coinductively as follows:
\begin{eqnarray*}
\merge{x::\sigma}{\tau} &\eqdef& x::\merge\tau\sigma\\
\split{x::y::\sigma} &\eqdef& (x::\split\sigma_1,\,y::\split\sigma_2).
\end{eqnarray*}
These functions exist and are unique, since they are the unique fixpoints of the eventually contractive maps
\begin{align*}
\alpha : (\SS^2\imp\SS)\ &\imp\ (\SS^2\imp\SS) &
\beta : (\SS\imp\SS^2)\ &\imp\ (\SS\imp\SS^2)
\end{align*}
defined by
\begin{eqnarray*}
\alpha(M)(x::\sigma,\,\tau) &\eqdef& x::M(\tau,\,\sigma)\\
\beta(S)(x::y::\sigma) &\eqdef& (x::S(\sigma)_1,\,y::S(\sigma)_2).
\end{eqnarray*}

We would like to show that $\mergename$ and $\splitname$ are inverses.  Traditionally, one would do this by exhibiting a bisimulation between $\mergename\,(\split{\sigma})$ and $\sigma$, thus concluding that $\mergename\,(\split{\sigma})=\sigma$, and another bisimulation between $\splitname\,(\merge{\sigma}{\tau})$ and $(\sigma,\,\tau)$, thus concluding that $\splitname\,(\merge{\sigma}{\tau})=(\sigma,\,\tau)$.

Here is how we would prove this result using the metric coinduction rule \eqref{eqn:coind1}.  Let $M:\SS^2\imp\SS$ and $S:\SS\imp\SS^2$.  If $M$ is a left inverse of $S$, then $\alpha^2(M)$ is a left inverse of $\beta(S)$:
\begin{eqnarray*}
\alpha^2(M)(\beta(S)(x::y::\sigma))
&=& \alpha(\alpha(M))(x::S(\sigma)_1,\,y::S(\sigma)_2)\\
&=& x::\alpha(M)(y::S(\sigma)_2,\,S(\sigma)_1)\\
&=& x::y::M(S(\sigma)_1,\,S(\sigma)_2)\\
&=& x::y::M(S(\sigma))\\
&=& x::y::\sigma.
\end{eqnarray*}
Similarly, if $M$ is a right inverse of $S$, then $\alpha^2(M)$ is a right inverse of $\beta(S)$:
\begin{eqnarray*}
\beta(S)(\alpha^2(M)(x::\sigma,\,y::\tau))
&=& \beta(S)(\alpha(\alpha(M))(x::\sigma,\,y::\tau))\\
&=& \beta(S)(x::\alpha(M)(y::\tau,\,\sigma))\\
&=& \beta(S)(x::y::M(\sigma,\,\tau))\\
&=& (x::S(M(\sigma,\,\tau))_1,\,y::S(M(\sigma,\,\tau))_2)\\
&=& (x::(\sigma,\,\tau)_1,\,y::(\sigma,\,\tau)_2)\\
&=& (x::\sigma,\,y::\tau).
\end{eqnarray*}
We conclude that if $M$ and $S$ are inverses, then so are $\alpha^2(M)$ and $\beta(S)$.

The property
\begin{eqnarray}
\phi(M,S) &\Iffdef& \mbox{$M$ and $S$ are inverses}\label{eqn:streamphi}
\end{eqnarray}
is a nonempty closed property of $(\SS^2\imp\SS)\times(\SS\imp\SS^2)$ which, as we have just shown, is preserved by the contractive map $(M,S)\mapsto(\alpha^2(M),\beta(S))$.  By \eqref{eqn:coind1}, $\phi$ holds of the unique fixpoint $(\mergename,\,\splitname)$.

That $\phi$ is nonempty and closed requires an argument, but these conditions typically follow from general topological considerations.  For example, \eqref{eqn:streamphi} is nonempty because the spaces $\SS$ and $\SS^2$ are both homeomorphic to the topological product of countably many copies of the discrete space $\Sigma$.

\section{Markov Chains}
\label{sec:mc}

A finite Markov chain is a finite state space, say $\{1,\ldots,n\}$, together with a stochastic matrix $P\in\reals^{n\times n}$ of transition probabilities, with $P_{st}$ representing the probability of a transition from state $s$ to state $t$ in one step.  The value $P^m_{st}$ is the probability that the system is in state $t$ after $m$ steps, starting in state $s$.

A fundamental result in the theory of Markov chains is that if $P$ is irreducible and aperiodic (definitions given below), then $P^m_{st}$ tends to $1/\mu_t$ as $m\imp\infty$, where $\mu_t$ is the \emph{mean first recurrence time} of state $t$, the expected time of first reentry into state $t$ after leaving state $t$.  Intuitively, if we expect to be in state $t$ about every $\mu_t$ steps, then in the long run we expect to be in state $t$ about $1/\mu_t$ of the time.

The proof of this result as given in Feller \cite{Fel} is rather lengthy, involving a complicated argument to establish the uniform convergence of a certain countable sequence of countable sequences.  The complete proof runs to several pages.  Introductory texts devote entire chapters to it (e.g.~\cite{Haggstrom02}) or omit the proof entirely (e.g.~\cite{MotwaniRaghavan95}).  In this section we show that, assuming some basic spectral properties of stochastic matrices, the coinduction rule can be used to give a simpler alternative proof.

\subsection{Spectral Properties}

Recall that $P$ is \emph{irreducible} if its underlying support graph is strongly connected.  The \emph{support graph} has vertices $\{1,\ldots,n\}$ and directed edges $\set{(s,t)}{P_{st}>0}$.  A directed graph is \emph{strongly connected} if there is a directed path from any vertex to any other vertex.  The matrix $P$ is \emph{aperiodic} if in addition, the gcd of the set $\set m{P^m_{ss}>0}$ is $1$ for all states $s$.  By the Perron--Frobenius theorem (see \cite{Bremaud99,Minc88}), if $P$ is irreducible and aperiodic, then $P$ has eigenvalue $1$ with multiplicity $1$ and all other eigenvalues have norm strictly less than $1$.

The matrix $P$ is itself not contractive, since $1$ is an eigenvalue.  However, consider the matrix
\begin{eqnarray*}
\sP,
\end{eqnarray*}
where $\one$ is the column vector of all $1$'s and $\xpose$ denotes matrix transpose.  The matrix $\frac 1n\one\one\xpose$ is the $n\times n$ matrix all of whose entries are $1/n$.

The spectra of $P$ and $\sP$ are closely related, as shown in the following lemma.

\begin{lem}
\label{lem:spec}
Let $P\in\reals^{n\times n}$ be a stochastic matrix.  Any (left) eigenvector $x\xpose$ of $\sP$ that lies in the hyperplane $x\xpose\one=0$ is also an eigenvector of $P$ with the same eigenvalue, and vice-versa.  The only other eigenvalue of $P$ is $1$ and the only other eigenvalue of $\sP$ is $0$.
\end{lem}
\proof
For any eigenvalue $\lambda$ of $P$ and corresponding eigenvector $x\xpose$,
\begin{eqnarray*}
\lambda x\xpose\one &=& x\xpose P\one = x\xpose\one
\end{eqnarray*}
since $P\one=\one$, so either $\lambda=1$ or $x\xpose\one=0$.  Similarly, for any eigenvalue $\lambda$ of $\sP$ and corresponding eigenvector $x\xpose$,
\begin{eqnarray*}
\lambda x\xpose\one &=& x\xpose(\sP)\one = x\xpose\one - x\xpose\one = 0,
\end{eqnarray*}
so either $\lambda=0$ or $x\xpose\one=0$.  But if $x\xpose\one=0$, then
\begin{eqnarray*}
x\xpose(P - \frac 1n\one\one\xpose) = x\xpose P - \frac 1n x\xpose\one\one\xpose = x\xpose P,
\end{eqnarray*}
so in this case $x\xpose$ is an eigenvector of $P$ iff it is an eigenvector of $\sP$ with the same eigenvalue.
\qed

\subsection{Coinduction and the Convergence of \texorpdfstring{$P^m$}{P^m}}
\label{sec:conv}

If $P$ is irreducible and aperiodic, then \mbox{$\sP$} is eventually contractive, since $\inf_n \sqrt[n]{\norm{(\sP)^n}}$ is equal to the \emph{spectral radius} or norm of the largest eigenvalue of $\sP$ (see \cite{DunfordSchwartz57}), which by Lemma \ref{lem:spec} is less than 1.  Thus the map
\begin{eqnarray}
x\xpose &\mapsto& x\xpose(\sP) + \frac 1n\one\xpose\label{eqn:markov}
\end{eqnarray}
is of the proper form to be used with the metric coinduction rule \eqref{eqn:coind1} to establish the convergence of $P^m$.

Since $\sP$ is eventually contractive, the map \eqref{eqn:markov} has a unique fixpoint $u\xpose$.  The set of stochastic vectors
\begin{eqnarray*}
S &=& \set{x\xpose}{x\xpose \geq 0,\ x\xpose\one=1}
\end{eqnarray*}
is closed and preserved by the map \eqref{eqn:markov}, since
\begin{eqnarray*}
x\xpose\one = 1 &\Imp& x\xpose(\sP) + \frac 1n\one\xpose = x\xpose P,\label{eqn:fixpt}
\end{eqnarray*}
and $S$ is preserved by $P$.  By the metric coinduction rule \eqref{eqn:coind1}, the unique fixpoint $u\xpose$ is contained in $S$.  By Lemma \ref{lem:spec}, it is also an eigenvector of $1$, and $y\xpose P^m$ tends to $u\xpose$ for any $y\xpose\in S$.  Applying this to the rows of any stochastic matrix $E$, we have that $EP^m$ converges to the matrix $\one u\xpose$.

\subsection{Recurrence Statistics}

Once we have established the convergence of $P^m$, we can give a much shorter argument than those of \cite{Fel,Haggstrom02} that the actual limit of $P^m_{st}$ is $1/\mu_t$.  We follow the notation of \cite{Fel}.

Fix a state $t$, and let $\mu=\mu_t$.  Let $f_m$ be the probability that after leaving state $t$, the system first returns to state $t$ at time $m$.  Let $u_m = P^m_{tt}$ be the probability that the system is in state $t$ at time $m$ after starting in state $t$.  By irreducibility, $\sum_{m=1}^\infty f_m = 1$ and $\mu = \sum_{m=1}^\infty mf_m <\infty$.  Let $\rho_m \eqdef \sum_{k=m+1}^\infty f_k$, and consider the generating functions
\begin{align*}
f(x) &\eqdef \sum_{m=1}^\infty f_mx^m & u(x) &\eqdef \sum_{m=0}^\infty u_mx^m\\
\rho(x) &\eqdef \sum_{m=0}^\infty \rho_mx^m & \sigma(x) &\eqdef u_0 + \sum_{m=0}^\infty (u_{m+1}-u_m)x^{m+1}.
\end{align*}
The probabilities $u_n$ obey the recurrence
\begin{align*}
u_0 &= 1 & u_n &= \sum_{m=0}^{n-1} u_mf_{n-m},
\end{align*}
which implies that $f(x)u(x) = u(x) - 1$. Elementary algebraic reasoning gives
\begin{eqnarray}
\sigma(x)\rho(x) &=& 1.\label{eqn:recip}
\end{eqnarray}
Now we claim that both $\sigma(1)$ and $\rho(1)$ converge.  The sequence $\rho(1)$ converges to $\mu > 0$, since
\begin{eqnarray}
\rho(1) &=& \sum_{m=1}^\infty \rho_m = \sum_{m=1}^\infty mf_m = \mu,\label{eqn:converge}
\end{eqnarray}
and the latter sequence in \eqref{eqn:converge} converges absolutely.  For $\sigma(1)$, we have
\begin{eqnarray*}
\sigma(1) &=& u_0 + \sum_{m=0}^\infty (u_{m+1} - u_m),
\end{eqnarray*}
which converges by the results of Section \ref{sec:conv}.  By \eqref{eqn:recip}, $\sigma(1)\rho(1)=1$, therefore $\sigma(1)=1/\mu$.  But the $m$th partial sum of $\sigma(1)$ is just $u_0 + \sum_{k=0}^{m-1} (u_{k+1} - u_k) = u_m$, so the sequence $u_m$ converges to $1/\mu$.

\section{Markov Decision Processes}
\label{sec:mdp}

In this section, we rederive some fundamental results on Markov decision processes using the metric coinduction principle.  A fairly general treatment of this theory is given in \cite{Denardo67}, and we follow the notation of that paper.  However, the strategic use of metric coinduction allows a more streamlined presentation.

\subsection{Existence of Optimal Strategies}

Let $V$ be the space of bounded real-valued functions on a set of states $\Omega$ with the sup norm $\norm v\eqdef\sup_{x\in\Omega} \len{v(x)}$.  The space $V$ is a complete metric space with metric $\norm{v-u}$.

For each state $x\in\Omega$, say we have a set $\Delta_x$ of \emph{actions}.  A \emph{deterministic strategy} is an element of $\Delta\eqdef\prod_{x\in\Omega} \Delta_x$, thus a selection of actions, one for each state $x\in\Omega$.  More generally, if $\Delta_x$ is a measurable space, let $\Meas{\Delta_x}$ denote the space of probability measures on $\Delta_x$.  A \emph{probabilistic strategy} is an element of $\prod_{x\in\Omega}\Meas{\Delta_x}$, thus a selection of probability measures, one for each $x\in\Omega$.  A deterministic strategy can be viewed as a probabilistic strategy in which all the measures are point masses.

Now suppose we have a \emph{utility function}
$h:\prod_{x\in\Omega}(\Delta_x\imp V\imp\reals)$ with the three properties listed below.\footnote{We write $h(x,\delta_x,u)$ instead of $h(x)(\delta_x)(u)$ for readability.}
The function $h$ induces a function $H$ such that $H_{\delta}(u)(x)=h(x,\delta_x,u)\in\reals$, where $x\in\Omega$, $\delta\in\Delta$, and $u\in V$.
\begin{enumerate}[(i)]
%\romanize
\item
The function $H$ is uniformly bounded as a function of $\delta$ and $x$.  That is, $H_\delta:V\imp V$, and for any fixed $u\in V$, $\sup_{\delta\in\Delta} \norm{H_\delta(u)}$ is finite.
\item
The functions $H_\delta$ are uniformly contractive with constant of contraction $c < 1$.  That is, for all $\delta\in\Delta$ and $u,v\in V$, $\norm{H_\delta(v)-H_\delta(u)}\leq c\cdot\norm{v-u}$.  Thus $H_\delta$ has a unique fixpoint, which we denote by $v_\delta$.
\item
Every $H_\delta$ is \emph{monotone}: if $u\leq v$, then $H_\delta(u)\leq H_\delta(v)$.  The order $\leq$ on $V$ is the pointwise order.
\end{enumerate}

\begin{lem}
Define $A:V\imp V$ by $A(u)(x)\eqdef\sup_{d\in\Delta_x} h(x,d,u)$.  The supremum exists since the $H_\delta$ are uniformly bounded.  Then $A$ is contractive with constant of contraction $c$.
\end{lem}
\proof
Let $\eps > 0$.  For $x\in\Omega$, assuming without loss of generality that $A(v)(x)\geq A(u)(x)$,
\reasoning{\eps + h(x,d,v) - \sup_{e\in\Delta_x} h(x,e,u)}
\begin{eqnarray*}
\lefteqn{\len{A(v)(x) - A(u)(x)}}\\
&=& \sup_{d\in\Delta_x} h(x,d,v) - \sup_{e\in\Delta_x} h(x,e,u)\\
&\leq& \reason{\eps + h(x,d,v) - \sup_{e\in\Delta_x} h(x,e,u)}{for suitably chosen $d\in\Delta_x$}\\
&\leq& \eps + h(x,d,v) - h(x,d,u)\\
&\leq& \eps + c\cdot\norm{v-u}.
\end{eqnarray*}
Since $\eps$ was arbitrary, $\len{A(v)(x) - A(u)(x)} \leq c\cdot\norm{v-u}$, thus
\begin{eqnarray*}
\norm{A(v) - A(u)} &\leq& \sup_x \len{A(v)(x) - A(u)(x)} \leq c\cdot\norm{v-u}.
\end{eqnarray*}
\qed

Since $A$ is contractive, it has a unique fixpoint $v\star$.

\begin{lem}
For any $\delta$, $v_\delta \leq v\star$.
\end{lem}
\proof
By the coinduction principle, it suffices to show that $u\leq v$ implies $H_\delta(u)\leq A(v)$.  Here the metric space is $V^2$, the closed property $\phi$ is $u\leq v$, and the contractive map is $(H_\delta,A)$.  But if $u\leq v$, then by monotonicity,
\begin{eqnarray*}
H_\delta(u)(x) &\leq& H_\delta(v)(x) = h(x,\delta_x,v) \leq \sup_{d\in\Delta_x} h(x,d,v) = A(v).
\end{eqnarray*}
\qed

\begin{lem}
The fixpoint $v\star$ can be approximated arbitrarily closely by $v_\delta$ for deterministic strategies $\delta$.
\end{lem}
\proof
Let $\eps > 0$.  Let $\delta$ be such that for all $x$,
\begin{eqnarray*}
\sup_{d\in\Delta_x} h(x,d,v\star) - h(x,\delta_x,v\star) &<& (1-c)\eps.
\end{eqnarray*}
We will show that $\norm{v\star - v_\delta} \leq \eps$.  By the coinduction rule \eqref{eqn:coind1}, it suffices to show that $\norm{v\star - u} \leq \eps$ implies $\norm{v\star - H_\delta(u)}\leq\eps$.  Here the metric space is $V$, the closed property $\phi(u)$ is $\norm{v\star - u} \leq \eps$, and the contractive map is $H_\delta$.  But if $\norm{v\star - u} \leq \eps$,
\begin{eqnarray*}
\norm{v\star - H_\delta(u)} &=& \sup_x\len{v\star(x) - H_\delta(u)(x)} = \sup_x\len{A(v\star)(x) - H_\delta(u)(x)}\\
&=& \sup_x\len{\sup_{d\in\Delta_x} h(x,d,v\star) - h(x,\delta_x,u)}\\
&\leq& \sup_x(\len{\sup_{d\in\Delta_x} h(x,d,v\star) - h(x,\delta_x,v\star)} + \len{h(x,\delta_x,v\star) - h(x,\delta_x,u)})\\
&\leq& (1-c)\eps + c\cdot\norm{v\star - u} \leq (1-c)\eps + c\eps = \eps.
\end{eqnarray*}
\qed

\subsection{Probabilistic Strategies}
\label{sec:prob}

We use the metric coinduction rule \eqref{eqn:coind1} to prove the well-known result that for Markov decision processes, probabilistic strategies are no better than deterministic strategies.  If $\sup_{d\in\Delta_x} h(x,d,v\star)$ is attainable for all $x$, then the deterministic strategy $\delta_x\eqdef\argmax_{d\in\Delta_x} h(x,d,v\star)$ is optimal, even allowing probabilistic strategies.  However, if $\sup_{d\in\Delta_x} h(x,d,v\star)$ is not attainable, then it is not so obvious what to do. 

For this argument, we assume that $\Delta_x$ is a measurable space and that for all fixed $x$ and $u$, $h(x,d,u)$ is an integrable function of $d\in\Delta_x$.  Given a probabilistic strategy $\mu:\prod_{x\in\Omega}\Meas{\Delta_x}$, the one-step utility function is $H_\mu:V\imp V$ defined by the Lebesgue integral
\begin{eqnarray*}
H_\mu(u)(x) &\eqdef& \int_{d\in\Delta_x} h(x,d,u)\cdot\mu_x(\triangle d).
\end{eqnarray*}
This integral accumulates the various individual payoffs over all choices of $d$ weighted by the measure $\mu_x$.

The map $H_\mu(u)$ is uniformly bounded in $\mu$, since
\begin{eqnarray*}
\norm{H_\mu(u)} &=& \sup_x\left|\int_{d\in\Delta_x} h(x,d,u)\cdot\mu_x(\triangle d)\right| \leq \sup_x\int_{d\in\Delta_x} \len{h(x,d,u)}\cdot\mu_x(\triangle d)\\
&\leq& \sup_x\sup_d\len{h(x,d,u)}\cdot \int_{d\in\Delta_x} \mu_x(\triangle d) = \sup_{x,d}\len{h(x,d,u)}.
\end{eqnarray*}
It is also a contractive map with constant of contraction $c$, since
\begin{eqnarray*}
\norm{H_\mu(v)-H_\mu(u)}
&=& \sup_x \len{H_\mu(v)(x)-H_\mu(u)(x)}\\
&=& \sup_x \left|\int_{d\in\Delta_x} h(x,d,v)\cdot\mu_x(\triangle d)-\int_{d\in\Delta_x} h(x,d,u)\cdot\mu_x(\triangle d)\right|\\
&=& \sup_x \left|\int_{d\in\Delta_x} (h(x,d,v)-h(x,d,u))\cdot\mu_x(\triangle d)\right|\\
&\leq& \sup_x \int_{d\in\Delta_x} \len{h(x,d,v)-h(x,d,u)}\cdot\mu_x(\triangle d)\\
&\leq& \sup_x \int_{d\in\Delta_x} c\cdot\norm{v-u}\cdot\mu_x(\triangle d)\\
&=& c\cdot\norm{v-u}\cdot\sup_x \int_{d\in\Delta_x} \mu_x(\triangle d)\\
&=& c\cdot\norm{v-u}.
\end{eqnarray*}
Since it is a contractive map, it has a unique fixpoint $v_\mu$.

Now take any deterministic strategy $\delta$ such that $h(x,\delta_x,v_\mu) \geq v_\mu(x)$ for all $x$.  This is always possible, since if $h(x,d,v_\mu) < v_\mu(x)$ for all $d\in\Delta_x$, then 
\begin{eqnarray*}
v_\mu(x) &=& H_\mu(v_\mu)(x) = \int_{d\in\Delta_x} h(x,d,v_\mu)\cdot\mu_x(\triangle d) < v_\mu(x),
\end{eqnarray*}
a contradiction.  The following lemma says that the deterministic strategy $\delta$ is no worse than the probabilistic strategy $\mu$.
\begin{lem}
$v_\delta \geq v_\mu$.
\end{lem}
\proof
Assuming $v_\mu\leq v$, we have
\begin{eqnarray*}
v_\mu(x) &\leq& h(x,\delta_x,v_\mu) \leq h(x,\delta_x,v) = H_\delta(v)(x),
\end{eqnarray*}
the second inequality by monotonicity.  As $x$ was arbitrary, $v_\mu \leq H_\delta(v)$.  The result follows from the coinduction principle on the metric space $V$ with $\phi(v)$ the closed property $v_\mu\leq v$ and contractive map $H_\delta$.
\qed

\section{Non-Well-Founded Sets}
\label{sec:nwf}

In classical Zermelo--Fraenkel set theory with choice (ZFC), the ``element of'' relation $\in$ is well-founded, as guaranteed by the axiom of foundation.  Aczel \cite{Aczel88} developed the theory of \emph{non-well-founded sets}, in which sets with infinitely descending $\in$-chains are permitted in addition to the well-founded sets.  These are precisely the sets that are explicitly ruled out of existence by the axiom of foundation.

In the theory of non-well-founded sets, the sets are represented by \emph{accessible pointed graphs} (\APG s).  An \APG\ is a directed graph with a distinguished node such that every node is reachable by a directed path from the distinguished node.  Two \APG s represent the same set iff they are bisimilar.  The \APG s of well-founded sets may be infinite, but may contain no infinite paths or cycles, whereas the \APG s of non-well-founded sets may contain infinite paths and cycles.  Equality as bisimulation is the natural analog of extensionality in ZFC; essentially, two \APG s are declared equal as sets if there is no witness among their descendants that forces them not to be.  The class $V$ is the class of sets defined in this way.

Aczel \cite{Aczel88} (see also \cite{BarwiseMoss96,Turi:thesis}) notes the strong role that coinduction plays in this theory.  Since equality between \APG s is defined in terms of bisimulation, coinduction becomes a primary proof technique for establishing the equivalence of different \APG s representing the same set.

In attempting to define a metric on non-well-founded sets, the classical Hausdorff distance suggests itself as a promising candidate.  This metric has been previously defined for the hereditarily finite well-founded sets and their completion, the \emph{finitary} non-well-founded sets, by Abramsky \cite{Abramsky05}.  For the more general case of arbitrary non-well-founded sets, there are two complications.  One is that we must apply the definition coinductively.  Another is that ordinarily, the Hausdorff metric is only defined on compact sets, since otherwise a Hausdorff distance of zero may not imply equality, and that is the case here.  However, the definition still makes sense even for non-compact sets and leads to further insights into the structure of non-well-founded sets.

In this section, we define a distance function $d:V^2\imp\reals$ based on a coinductive application of the Hausdorff distance function and derive some properties of $d$.  We show that $(V,d)$ forms a compact pseudometric space.  Being a pseudometric instead of a metric means that there are sets $s\neq t$ with $d(s,t)=0$.  Nevertheless, we identify a maximal family of sets that includes all the hereditarily finite sets on which $d$ acts as a metric.

We will prove the following results.  Define $s\approx t$ if $d(s,t)=0$.  Call a set $s$ \emph{singular} if the only $t$ such that $s\approx t$ is $s$ itself.
\begin{enumerate}[$\bullet$]
\item
A set is singular if and only if it is hereditarily finite.
\item
All singular sets are closed in the pseudometric topology.  In particular, all hereditarily finite sets are hereditarily closed (but not vice-versa).
\item
A set is hereditarily closed if and only if it is closed and all elements are singular.  
\item
All hereditarily closed sets are canonical (but not vice-versa), where a set is \emph{canonical} if it is a member of a certain coinductively-defined class of canonical representatives of the $\approx$-classes.
\item
The map $d$ is a metric on the canonical sets; moreover, the canonical sets are a maximal class for which this is true.
\end{enumerate}

\newcommand\PP{\mathcal P}
\subsection{Coinductive Definition of Functions}

Just as classical ZFC allows the definition of functions by induction over ordinary well-founded sets, there is a corresponding principle for non-well-founded sets known as the \emph{Solution Lemma} \cite{Aczel88,Turi:thesis}.  In particular, the Solution Lemma implies that for any function $H:V\imp V$, the equation
\begin{eqnarray}
G(s) &\eqdef& \set{G(u)}{u\in H(s)}\label{eqn:nwf11}
\end{eqnarray}
determines $G:V\imp V$ uniquely.  This is because if $G$ and $G'$ both satisfy \eqref{eqn:nwf11}, then the relation
\begin{eqnarray*}
u\mathrel R v &\Iffdef& \exists s\ u=G(s)\wedge v=G'(s)
\end{eqnarray*}
is a bisimulation, therefore $G(s)=G'(s)$ for all $s$.  In coalgebraic terms\footnote{When regarding $V$ as a coalgebra, the notation $(V,\in)$ is a slight but convenient abuse.  Formally, these structures are coalgebras with respect to the powerset functor $\PP$.  To be precise, we should write $(V,\,\beta)$, where $\beta:V\imp\PP V$ and write $s\in\beta(t)$ instead of $s\in t$.}, the map $G$ is the unique morphism from the coalgebra $(V,\set{(s,t)}{s\in H(t)})$ to the final coalgebra $(V,\in)$; see \cite[Chp.~7]{Aczel88} or \cite[Part~V]{Turi:thesis}.

\subsection{Definition of \texorpdfstring{$d$}{d}}

Let $B$ be the Banach space of bounded real-valued functions $g:\APG^2\imp\reals$ with norm
\begin{eqnarray*}
\norm g &\eqdef& \sup_{s,t} \len{g(s,t)}.
\end{eqnarray*}
Define the map $\tau: B\imp B$ by
\begin{eqnarray*}
\tau(g)(s,t) &\eqdef& \left\{\begin{array}{ll}
0 & \mbox{if $s,t=\emptyset$}\\[1ex]
1 & \mbox{if $s=\emptyset \Iff t\neq\emptyset$}\\
\frac 12\max\left\{\begin{array}{l}
\sup_{u\in s}\inf_{v\in t} g(u,v)\\
\sup_{v\in t}\inf_{u\in s} g(u,v)
\end{array}\right.
& \mbox{if $s,t\neq\emptyset$.}
\end{array}\right.
\end{eqnarray*}
It can be shown that $\norm{\tau(g) - \tau(g')} \leq \frac 12\norm{g-g'}$, thus $\tau$ is contractive on $B$ with constant of contraction $1/2$ and has a unique fixpoint $d\in B$.  One can therefore use the metric coinduction rule \eqref{eqn:coind1} to prove properties of $d$.

To illustrate, let us show that the non-well-founded sets $V$ form a compact (thus complete) pseudometric space with respect to the distance function $d$.  At the outset, it is not immediately clear that $d$ is well-defined on $V$.  We must argue that $d$ is invariant on bisimulation classes; that is, for any bisimulation $\mathrel R$, if $s \mathrel R s'$ and $t \mathrel R t'$, then $d(s,t)=d(s',t')$.  We will use the metric coinduction rule \eqref{eqn:coind1} to prove this.

Consider the following closed property on $B$, defined with respect to an arbitrary but fixed bisimulation $\mathrel R$ on the class of \APG s:
\begin{eqnarray*}
\phi(g) &\Iffdef& \forall s\ \forall s'\ \forall t\ \forall t'\ s \mathrel R s' \wedge t \mathrel R t'\ \Imp\ g(s,t)=g(s',t').
\end{eqnarray*}
This property is closed in the metric topology on $B$, since it is an infinite conjunction of closed properties $g(s,t)=g(s',t')$, one for each selection of $s,s',t,t'$ such that $s \mathrel R s'$ and $t \mathrel R t'$.  It is clearly nonempty.  We wish to prove that $\phi(d)$.  By the metric coinduction rule \eqref{eqn:coind1}, it suffices to show that $\phi$ is closed under $\tau$.

Suppose $\phi(g)$.  We want to show that $\phi(\tau(g))$, or in other words,
\begin{eqnarray*}
\forall s\ \forall s'\ \forall t\ \forall t'\ s \mathrel R s' \wedge t \mathrel R t' &\Imp& \tau(g)(s,t)=\tau(g)(s',t').
\end{eqnarray*}
Let $s,s',t,t'$ be such that $s \mathrel R s'$ and $t \mathrel R t'$.  Since $\mathrel R$ is a bisimulation, we have
\begin{eqnarray*}
\forall u\in s\ \exists u'\in s'\ u\mathrel R u' &\quad& \forall u'\in s'\ \exists u\in s\ u\mathrel R u'\\
\forall v\in t\ \exists v'\in t'\ v\mathrel R v' &\quad& \forall v'\in t'\ \exists v\in t\ v\mathrel R v'.
\end{eqnarray*}
It follows that $s=\emptyset$ iff $s'=\emptyset$ and $t=\emptyset$ iff $t'=\emptyset$.  If $s=s'=\emptyset$, then 
\begin{eqnarray*}
\tau(g)(s,t) &=& \left\{\begin{array}{ll}
0 & \mbox{if $t,t'=\emptyset$}\\[1ex]
1 & \mbox{if $t,t'\neq\emptyset$}
\end{array}\right\}\ \ =\ \ \tau(g)(s',t').
\end{eqnarray*}
A symmetric argument holds if $t=t'=\emptyset$.

Otherwise, all four sets $s,s',t,t'$ are nonempty.  In this case,
\begin{eqnarray*}
\tau(g)(s,t) &=& 
\textstyle\frac 12\max\left\{\begin{array}{l}
\sup_{u\in s}\inf_{v\in t} g(u,v)\\
\sup_{v\in t}\inf_{u\in s} g(u,v)
\end{array}\right.\\
\tau(g)(s',t') &=& 
\textstyle\frac 12\max\left\{\begin{array}{l}
\sup_{u'\in s'}\inf_{v'\in t'} g(u',v')\\
\sup_{v'\in t'}\inf_{u'\in s'} g(u',v'),
\end{array}\right.
\end{eqnarray*}
so it suffices to show that
\begin{eqnarray}
\sup_{u\in s}\inf_{v\in t} g(u,v) &=& \sup_{u'\in s'}\inf_{v'\in t'} g(u',v')\label{eqn:nwf2a}\\
\sup_{v\in t}\inf_{u\in s} g(u,v) &=& \sup_{v'\in t'}\inf_{u'\in s'} g(u',v').\label{eqn:nwf2b}
\end{eqnarray}
We show only \eqref{eqn:nwf2a}; the argument for \eqref{eqn:nwf2b} is symmetric.  Also by symmetry, we need only show the inequality in one direction:
\begin{eqnarray*}
\sup_{u\in s}\inf_{v\in t} g(u,v) &\leq& \sup_{u'\in s'}\inf_{v'\in t'} g(u',v').
\end{eqnarray*}
This inequality follows from the property
\[
\forall u\in s\ \exists u'\in s'\ \inf_{v\in t} g(u,v) \leq \inf_{v'\in t'} g(u',v'),
\]
which in turn follows from
\[
\forall u\in s\ \exists u'\in s'\ \forall v'\in t'\ \exists v\in t\ g(u,v) \leq g(u',v').
\]
In fact, we have
\[
\forall u\in s\ \exists u'\in s'\ \forall v'\in t'\ \exists v\in t\ g(u,v) = g(u',v')
\]
by choosing $u'\in s'$ such that $u\mathrel R u'$ and $v\in t$ such that $v\mathrel R v'$, as guaranteed by the coinduction hypothesis and the fact that $\mathrel R$ is a bisimulation.

We conclude by the metric coinduction principle \eqref{eqn:coind1} that $\phi(d)$ holds, thus $d$ is invariant on the equivalence classes of any bisimulation $\mathrel R$ on \APG s, therefore well-defined on $V$.

To show that $d$ is a pseudometric, we must also show
\begin{align*}
d(s,t) &\geq 0\ \mbox{(in fact, $d(s,t)\in [0,1]$)} & d(s,t) &= d(t,s)\\
d(s,u) &\leq d(s,t) + d(t,u) & d(s,s) &= 0.
\end{align*}
All these properties can be shown in the same way, by metric coinduction.  One need only argue that they are all nonempty closed properties closed under $\tau$.

We will establish compactness (hence completeness) later in section \ref{sec:compactness}, but first we introduce the canonical sets.

\renewcommand\bar[1]{\overline{#1}}
\subsection{Canonical Sets}

\begin{figure}[ht]
\begin{center}
\begin{picture}(0,110)(-60,-100)
\put(0,0){\cir}
\put(0,0){\line(-3,-1){60}}
\put(0,0){\line(-2,-1){40}}
\put(0,0){\line(-1,-1){20}}
\put(0,0){\line(0,-1){20}}
\put(0,0){\line(1,-1){20}}
\put(0,0){\line(2,-1){40}}
\put(0,0){\line(3,-1){60}}
\put(0,0){\line(5,-1){100}}
\multiput(-60,-20)(20,0)7{\cir}
\multiput(-40,-20)(20,0)6{\line(0,-1){10}}
\multiput(-40,-30)(20,0)6{\cir}
\multiput(-20,-30)(20,0)5{\line(0,-1){10}}
\multiput(-20,-40)(20,0)5{\cir}
\multiput(0,-40)(20,0)4{\line(0,-1){10}}
\multiput(0,-50)(20,0)4{\cir}
\multiput(20,-50)(20,0)3{\line(0,-1){10}}
\multiput(20,-60)(20,0)3{\cir}
\multiput(40,-60)(20,0)2{\line(0,-1){10}}
\multiput(40,-70)(20,0)2{\cir}
\multiput(60,-70)(20,0)1{\line(0,-1){10}}
\multiput(60,-80)(20,0)1{\cir}
\put(100,-20){\cir}
\put(100,-30){\circle{20}}
\put(70,-20){\makebox(0,0)[l]{$\cdots$}}
\put(70,-85){\makebox(0,0)[l]{$\ddots$}}
\put(20,-100){\makebox(0,0){$\set{\bar n}{n\geq 0}\union\Omega$}}
\end{picture}
\begin{picture}(0,110)(110,-100)
\put(0,0){\cir}
\put(0,0){\line(-3,-1){60}}
\put(0,0){\line(-2,-1){40}}
\put(0,0){\line(-1,-1){20}}
\put(0,0){\line(0,-1){20}}
\put(0,0){\line(1,-1){20}}
\put(0,0){\line(2,-1){40}}
\put(0,0){\line(3,-1){60}}
\multiput(-60,-20)(20,0)7{\cir}
\multiput(-40,-20)(20,0)6{\line(0,-1){10}}
\multiput(-40,-30)(20,0)6{\cir}
\multiput(-20,-30)(20,0)5{\line(0,-1){10}}
\multiput(-20,-40)(20,0)5{\cir}
\multiput(0,-40)(20,0)4{\line(0,-1){10}}
\multiput(0,-50)(20,0)4{\cir}
\multiput(20,-50)(20,0)3{\line(0,-1){10}}
\multiput(20,-60)(20,0)3{\cir}
\multiput(40,-60)(20,0)2{\line(0,-1){10}}
\multiput(40,-70)(20,0)2{\cir}
\multiput(60,-70)(20,0)1{\line(0,-1){10}}
\multiput(60,-80)(20,0)1{\cir}
\put(70,-20){\makebox(0,0)[l]{$\cdots$}}
\put(70,-85){\makebox(0,0)[l]{$\ddots$}}
\put(0,-100){\makebox(0,0){$\set{\bar n}{n\geq 0}$}}
\end{picture}
\label{fig:omegatree}
\end{center}
\caption{Distinct sets of distance $0$}
\label{fig:notsing}
\end{figure}
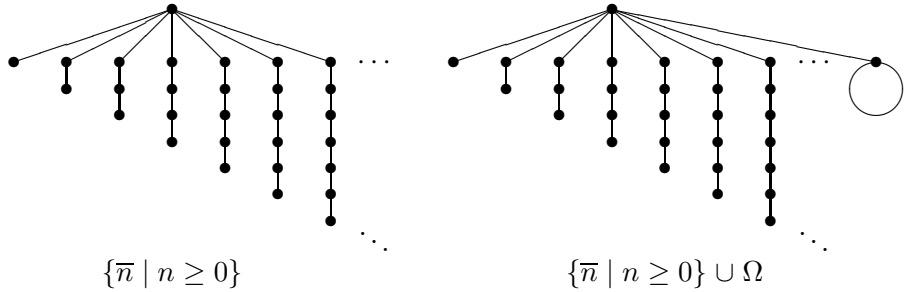

The map $d$ is only a pseudometric and not a metric, since it is possible that $d(s,t)=0$ even though $s\neq t$.  For example, define $\bar 0 = \emptyset$, $\bar{n+1}=\{\bar n\}$.  Let $\Omega$ be the unique non-well-founded set such that $\Omega=\{\Omega\}$.  The sets $\set{\bar n}{n\geq 0}$ and $\set{\bar n}{n\geq 0}\union\Omega$ are distinct, but distance $0$ apart (Fig.~\ref{fig:notsing}).  This follows from the observation that $d(\bar n,\Omega)=2^{-n}$, so
\begin{eqnarray*}
\sup_{v\in\set{\bar n}{n\geq 0}\union\Omega}\ \inf_{u\in\set{\bar n}{n\geq 0}} d(u,v)
&=& \inf_{{n\geq 0}} d(\bar n,\Omega)\ \ =\ \ 0.
\end{eqnarray*}
Nevertheless, it is possible to relate this map to the coalgebraic structure of $V$.  

The map $d$ defines a pseudometric topology with basic open neighborhoods $\set t{d(s,t) < \eps}$ for each set $s$ and $\eps>0$, but because $d$ is only a pseudometric, the topology does not have nice separation properties.  However, if we define $s\approx t \Iffdef d(s,t)=0$, then $d$ is well-defined on $\approx$-equivalence classes and is a metric on the quotient space.

More interestingly, we can identify a natural class of canonical elements, one in each $\approx$-class, such that $d$, restricted to canonical elements, is a metric; moreover, the canonical elements are a maximal class for which this is true.  Thus the quotient space is isometric to the subspace of canonical elements.  The canonical elements include all the hereditarily finite sets.  

The canonical elements are defined as the images of the function $F:V\imp V$, defined coinductively as follows:
\begin{eqnarray}
F(s) &\eqdef& \set{F(u)}{u\in\cl(s)},\label{eqn:nwf10}
\end{eqnarray}
where $\cl$ denotes closure in the pseudometric topology.  The equation \eqref{eqn:nwf10} determines $F$ uniquely, as with \eqref{eqn:nwf11}.  A set $s$ is called \emph{canonical} if $s=F(t)$ for some $t$; equivalently, by Corollary \ref{cor:F}(ii) below, if $s$ is a fixpoint of $F$.  For example, the right-hand side of Fig.~\ref{fig:notsing} is $F$ applied to the left-hand side, and the set on the right-hand side is canonical.

\begin{lem}
\label{lem:cl0}
$d(s,t)=0$ iff $\cl(s)=\cl(t)$.
\end{lem}
\proof
If $s=t=\emptyset$, then both sides are true.  If exactly one of $s,t$ is $\emptyset$, then both sides are false.  Finally, if both $s,t\neq\emptyset$, then
\begin{eqnarray*}
d(s,t)=0 &\Iff& \sup_{u\in s}\inf_{v\in t}d(u,v)=0 \wedge \sup_{v\in t}\inf_{u\in s}d(u,v)=0\\
&\Iff& \forall u\in s\ \forall\eps>0\ \exists v\in t\ d(u,v)<\eps
\wedge \forall v\in t\ \forall\eps>0\ \exists u\in s\ d(u,v)<\eps\\
&\Iff& s\subs\cl(t) \wedge t\subs\cl(s)\\
&\Iff& \cl(s)=\cl(t).
\end{eqnarray*}
\qed

\begin{thm}\hfil
\begin{enumerate}[\em(i)]
%\romanize
\item
If $d(s,t)=0$, then $F(s)=F(t)$.
\item
For all $s$, $d(s,F(s))=0$; that is, $s\approx F(s)$.
\end{enumerate}
\end{thm}
\proof\hfil
\begin{enumerate}[(i)]
\item
By Lemma \ref{lem:cl0}, if $d(s,t)=0$, then $\cl(s)=\cl(t)$, and the conclusion $F(s)=F(t)$ is immediate from \eqref{eqn:nwf10}.

\item
We proceed by coinduction on the definition of $d$.  We strengthen the coinduction hypothesis $g(s,F(s))=0$ with the two extra assertions that $0\leq g(s,t)\leq d(s,t)$ and that $g$ satisfies the triangle inequality.  We wish to show that this combined property holds of $\tau(g)$ under the assumption that it holds of $g$.

That $0\leq \tau(g)(s,t)\leq \tau(d)(s,t) = d(s,t)$ is clear from the coinduction hypothesis and the monotonicity of the operators in the definition of $\tau$.  The argument that $\tau(g)$ satisfies the triangle inequality is equally straightforward.  Thus it remains to show that $\tau(g)(s,F(s))=0$.

By definition of $F$, $s=\emptyset$ iff $F(s)=\emptyset$, and in this case $\tau(g)(s,F(s))=0$ by definition of $\tau$.  Otherwise $s\neq\emptyset$ and $F(s)\neq\emptyset$.  To show $\tau(g)(s,F(s))=0$ in this case, we need to show that
\begin{eqnarray*}
\sup_{u\in s} \inf_{v\in F(s)} g(u,v) = \sup_{u\in s} \inf_{w\in\cl(s)} g(u,F(w)) &=& 0,\\
\sup_{v\in F(s)} \inf_{u\in s} g(u,v) = \sup_{w\in\cl(s)} \inf_{u\in s} g(u,F(w)) &=& 0.
\end{eqnarray*}
It suffices to show
\begin{eqnarray*}
\forall u\in s\ \inf_{w\in\cl(s)} g(u,F(w)) = 0, &\quad& \forall w\in\cl(s)\ \inf_{u\in s} g(u,F(w)) = 0.
\end{eqnarray*}
For the former, we can take $w=u$; then the result follows from the coinduction hypothesis $g(u,F(u))=0$.  For the latter, let $w\in\cl(s)$.  Here we use all three clauses of the coinduction hypothesis:
\begin{eqnarray*}
\inf_{u\in s} g(u,F(w)) &\leq& \inf_{u\in s} g(u,w) + g(w,F(w)) \leq \inf_{u\in s} d(u,w) + 0 = 0,
\end{eqnarray*}
the last equation from the fact that $w\in\cl(s)$.
\qed
\end{enumerate}

\begin{cor}\hfil
\label{cor:F}
\begin{enumerate}[\em(i)]
%\romanize
\item
$d(s,t)=0$ iff $F(s)=F(t)$.
\item
For all $s$, $F(F(s))=F(s)$.
\item
Every $\approx$-equivalence class contains exactly one canonical set, and $d$ restricted to canonical sets is a metric.  Moreover, the canonical sets are a maximal class for which this is true.
\end{enumerate}
\end{cor}

\subsection{Compactness}
\label{sec:compactness}

For the results of section \ref{sec:nwfmain}, we need to show that the space of non-well-founded sets is compact under $d$, thus complete.  We will show that every infinite set has a limit point.  Define the equivalence relations $\approx_n$ inductively by:
\begin{align*}
s &\approx_0 t\ \mbox{for all $s,t$} & s \approx_{n+1} t &\Iffdef \parbox[t]{4cm}{$\forall u\in s\ \exists v\in t\ u\approx_n v\\
\wedge\ \forall v\in t\ \exists u\in s\ u\approx_n v$.}
\end{align*}
Also define inductively
\begin{align*}
S_0 &\eqdef \emptyset & S_{n+1} &\eqdef \powerset{S_n},
\end{align*}
where $\powerset{A}$ denotes the powerset of $A$.  Each $S_n$ is a well-founded hereditarily finite set.  For $n\geq 0$, define the map $f_n:V\imp S_{n+1}$ inductively by
\begin{align*}
f_0(s) &\eqdef \emptyset & f_{n+1}(s) &\eqdef \set{f_n(u)}{u\in s}.
\end{align*}

The following properties of $S_n$, $\approx_n$, and $f_n$ are easily established by induction on $n$.
\begin{lem}
\label{lem:flem1}
For all $s,t\in V$ and $m,n\geq 0$,
\begin{enumerate}[\em(i)]
%\romanize
\item
$f_n(s)\in S_{n+1}$;
\item
if $s\in S_{n+1}$ then $f_n(s)=s$;
\item
$s\approx_n f_n(s)$;
\item
$f_n(f_m(s))=f_{\min\,m,n}(s)$;
\item
if $s,t\in S_{n+1}$ and $s\approx_n t$, then $s=t$.
\end{enumerate}
\end{lem}

\begin{lem}
\label{lem:flem2}
For all $s,t\in V$ and $n\geq 0$, the following are equivalent:
\begin{enumerate}[\em(i)]
%\romanize
\item
$s\approx_n t$;
\item
$f_n(s)=f_n(t)$;
\item
$d(s,t)\leq 2^{-n}$.
\end{enumerate}
\end{lem}

For each $s\in V$, let $f(s)$ denote the sequence $f_0(s),f_1(s),f_2(s),\ldots$.  It follows from Lemma \ref{lem:flem1}(iv) that $f_n(f_{n+1}(s)) = f_n(s)$.  Moreover, we have the following representation theorem as converse:

\begin{lem}
\label{lem:rep}
Any sequence $s_0,s_1,s_2,\ldots$ such that $f_n(s_{n+1})=s_n$ for all $n\geq 0$ is $f(s)$ for some $s$.
\end{lem}
\proof
Let $W$ be the set of all sequences $s=s_0,s_1,s_2,\ldots$ such that $s_n=f_n(s_{n+1})$, $n\geq 0$.  This is a set, since the defining condition implies $s_n\in S_{n+1}$.  Consider the system with nodes $W$ and edges $\mathrel N$ defined by
\begin{eqnarray*}
u\mathrel N s &\Iffdef& \forall n\geq 0\ u_n\in s_{n+1}.
\end{eqnarray*}
We claim that $f_n(s)=s_n$.  The proof is by induction on $n$.  Certainly $f_0(s)=\emptyset=s_0$, since $s_0=f_0(s_1)\in S_1$ and $\emptyset$ is the only element of $S_1$.  Now suppose the claim is true for $n$.  Then
\begin{eqnarray*}
f_{n+1}(s) &=& \set{f_n(u)}{u\in W,\ u\mathrel N s} = \set{f_n(u)}{u\in W,\ \forall k\ u_k\in s_{k+1}}\\
&=& \set{u_n}{u\in W,\ \forall k\ u_k\in s_{k+1}} = s_{n+1}.
\end{eqnarray*}
The last equation requires that for all $a\in S_{n+1}$, there exists $u\in W$ such that $u_n=a$.  The sequence $u=f_0(a),f_1(a),f_2(a),\ldots$ does it.
\qed

\begin{lem}
\label{lem:compactness}
The space $V$ is compact under the pseudometric $d$, therefore complete.
\end{lem}
\proof
We wish to show that every infinite set $s$ has a limit point $p$ (not necessarily contained in $s$).  Let $W$ be the tree of all sequences $u_0,u_1,u_2,\ldots$ such that $f_n(u_{n+1})=u_n$ for all $n\geq 0$ as defined in the proof of Lemma \ref{lem:rep}.  This is a finitely branching, infinite tree with root $\emptyset$.  By K\"onig's lemma, there is an infinite path $p$ in $W$ such that for every node $p_n$ on the path, there 
are infinitely many $u\in s$ such that $f_n(u)=p_n$.  The set represented by the path $p$ as given by Lemma \ref{lem:rep} is the desired limit point, since for all $k$, there exist infinitely many $u\in s$ such that $f_k(u)=p_k=f_k(p)$, therefore $d(u,p)\leq 2^{-k}$.
\qed

\subsection{Hereditarily Finite Sets Are Canonical}
\label{sec:nwfmain}

Let $\phi$ be a property of sets.  We define a set to be \emph{hereditarily $\phi$} (\Hphi) if it has an \APG\ representation in which every node represents a set satisfying $\phi$.  Equivalently, \Hphi\ is the largest solution of
\begin{eqnarray*}
\Hphi(s) &\Iffdef& \phi(s) \wedge \forall u\in s\ \Hphi(u).
\end{eqnarray*}

The \emph{hereditarily finite} (\HF) sets are those possessing an \APG\ representation
in which every node has finite out-degree (not necessarily bounded).
\begin{figure}[ht]
%\begin{wrapfigure}{r}{10mm}
%\begin{flushright}
\begin{picture}(102,64)(-22,-64)
\thicklines
\multiput(0,0)(20,-20)3{\cir}
\multiput(0,0)(20,-20)3{\vector(1,-1){20}}
\multiput(0,0)(20,-20)3{\vector(-1,-1){20}}
\put(-22,-22){\makebox(0,0)[tr]{$\bar 0$}}
\put(-2,-42){\makebox(0,0)[tr]{$\bar 1$}}
\put(18,-62){\makebox(0,0)[tr]{$\bar 2$}}
\multiput(65,-65)(5,-5)3{\makebox(0,0)[tl]{$\cdot$}}
\end{picture}
%\end{flushright}
\caption{$f(0)$}
%\vspace{-6mm}
\label{fig:aczel}
%\end{wrapfigure}
\end{figure}Note that this differs from Aczel's definition \cite[p.~7]{Aczel88}.  Aczel defines a set to be hereditarily finite if it has a finite \APG, which is a much stronger condition.  Aczel's definition and ours coincide for well-founded sets by K\"onig's lemma, but not for non-well-founded sets in general.  For example, the set $f(0)$, where $f$ is defined coinductively by $f(n) = \{\bar n,f(n+1)\}$ (Fig.~\ref{fig:aczel}), is hereditarily finite in our sense but not Aczel's.
We would prefer the term \emph{regular} or \emph{rational} for sets that are hereditarily finite in Aczel's sense, since they are exactly the sets that have a regular or rational tree representation \cite{Courcelle83}.

A set is \emph{hereditarily closed} (\HC) if it has an \APG\ representation in which every node represents a closed set in the pseudometric topology.  Recall that a set is \emph{singular} if it forms a singleton $\approx$-class.

\begin{lem}
\label{lem:sing}
If $s$ is singular, then all elements of $s$ are singular.  Thus all singular sets are hereditarily singular.
\end{lem}
\proof
Suppose $u\in s$, $v\neq u$, and $d(u,v)=0$.  We claim that (i) if $v\not\in s$, then $d(s,s\union\{v\})=0$,
and (ii) if $v\in s$, then $d(s,s-\{v\})=0$, thus in either case, $s$ is not singular.

In case (i), we must show
\begin{eqnarray*}
\sup_{x\in s}\inf_{y\in s\union\{v\}} d(x,y) = 0, &\quad& \sup_{y\in s\union\{v\}}\inf_{x\in s} d(x,y) = 0.
\end{eqnarray*}
It suffices to show
\begin{eqnarray*}
\forall x\in s\ \exists y\in s\union\{v\}\ d(x,y) = 0, &\quad& \forall y\in s\union\{v\}\ \exists x\in s\ d(x,y) = 0.
\end{eqnarray*}
The former is immediate by picking $y=x$.  For the latter, pick $x=y$ if $y\neq v$, otherwise pick $x=u$.

Case (ii) is really the same case as (i), with $s-\{v\}$ in (ii) playing the role of $s$ in (i).
\qed

\begin{lem}
\label{lem:cl}\mbox{\ }
\begin{enumerate}[\em(i)]
%\romanize
\item
If $s$ is closed and all elements of $s$ are closed, then all elements of $s$ are singular.
\item
Every singular set is closed.
\end{enumerate}
\end{lem}
\proof\hfil
\begin{enumerate}[(i)]
\item
Suppose $u\in s$ and $d(u,v)=0$.  Then $v\in s$, since $s$ is closed.  By Lemma \ref{lem:cl0}, $\cl(u)=\cl(v)$.  But $u$ and $v$ are both closed, so $u=v$.

\item
By Lemma \ref{lem:cl0}, $d(\cl(u),u)=0$, so if $u$ singular then $u=\cl(u)$.
\qed
\end{enumerate}

\begin{thm}
\label{thm:cl}
A set is hereditarily closed if and only if it is closed and all its elements are singular.
\end{thm}
\proof
This follows directly from Lemmas \ref{lem:sing} and \ref{lem:cl}.
\qed

\begin{thm}
\label{thm:finsing}
A set is singular if and only if it is hereditarily finite.
\end{thm}
\proof
Suppose first that $s$ is hereditarily finite (\HF).  Consider the binary relation on sets $s,t$ defined by
\begin{eqnarray}
\HF(s)\wedge d(s,t)=0.\label{eqn:nwf1}
\end{eqnarray}
We have
\reasoning{\forall v\in t\ \exists u\in s\ \HF(u)\wedge d(u,v)=0}
\begin{eqnarray}
\HF(s)\wedge d(s,t)=0
&\Imp& \forall v\in t\ \forall\eps>0\ \exists u\in s\ \HF(u)\wedge d(u,v)<\eps\nonumber\\
&\Imp& \forall v\in t\ \exists u\in s\ \HF(u)\wedge d(u,v)=0,\label{eqn:nwf0}
\end{eqnarray}
since $u$ is finite.  It follows that
\begin{eqnarray*}
\HF(s)\wedge\HF(t)\wedge d(s,t)=0
&\Imp& \forall u\in s\ \exists v\in t\ \HF(u)\wedge\HF(v)\wedge d(u,v)=0\\
&& \wedge\ \forall v\in t\ \exists u\in s\ \HF(u)\wedge\HF(v)\wedge d(u,v)=0,
\end{eqnarray*}
so the binary relation $\HF(s)\wedge\HF(t)\wedge d(s,t)=0$ is a bisimulation; thus
\begin{eqnarray*}
\HF(s)\wedge\HF(t)\wedge d(s,t)=0 &\Imp& s=t.
\end{eqnarray*}
Thus if $\HF(s)$, then there is a positive lower bound $\delta>0$ on $d(u,v)$ for $u,v\in s$, $u\neq v$.  But then
\begin{eqnarray*}
\HF(s)\wedge d(s,t)=0
&\Imp& \forall u\in s\ \forall\eps>0\ \exists v\in t\ \HF(u)\wedge d(u,v)<\eps\\
&\Imp& \forall u\in s\ \exists v\in t\ \HF(u)\wedge d(u,v)<\delta,
\end{eqnarray*}
and using \eqref{eqn:nwf0}, this gives
\begin{eqnarray*}
\lefteqn{\HF(s)\wedge d(s,t)=0}\hspace{2em}\\
&\Imp& \forall u\in s\ \exists v\in t\ \HF(u)\wedge d(u,v)<\delta \wedge \exists w\in s\ d(w,v)=0\\
&\Imp& \forall u\in s\ \exists v\in t\ \exists w\in s\ \HF(u)\wedge d(w,v)=0\wedge d(u,w)<\delta\\
&\Imp& \forall u\in s\ \exists v\in t\ \exists w\in s\ \HF(u)\wedge d(w,v)=0\wedge u=w\\
&\Imp& \forall u\in s\ \exists v\in t\ \HF(u)\wedge d(u,v)=0.
\end{eqnarray*}
This combined with \eqref{eqn:nwf0} says that the relation \eqref{eqn:nwf1} itself is a bisimulation.  Thus $\HF(s) \wedge d(s,t)=0$ implies $s=t$; in other words, $\HF(s)$ implies that $s$ is singular.

Now suppose that $s$ is singular.  By Lemma \ref{lem:sing}, $s$ is hereditarily singular.  We argue that $s$ must be finite.  If $s$ is infinite, then by Lemma \ref{lem:compactness}, $s$ has a limit point $p$ (not necessarily contained in $s$).  We claim that (i) if $p\not\in s$, then $d(s,s\union\{p\})=0$, and (ii) if $p\in s$, then $d(s,s-\{p\})=0$, thus in either case $s$ is not singular.  For (i),
\begin{eqnarray*}
d(s,s\union\{p\})=0 &\Iff& \forall u\in s\ \forall\eps > 0\ \exists v\in s\union\{p\}\ d(u,v) < \eps\\
&& \wedge\ \forall v\in s\union\{p\}\ \forall\eps > 0\ \exists u\in s\ d(u,v) < \eps.
\end{eqnarray*}
The first clause is true by taking $v=u$.  For the second clause, we can take $u=v$ unless $v=p$.  But if $v=p$, the condition reduces to
\begin{eqnarray*}
&& \forall\eps > 0\ \exists u\in s\ d(u,p) < \eps,
\end{eqnarray*}
which is true by Lemma \ref{lem:flem2}.

Case (ii) is really the same as case (i), with $s-\{p\}$ in (ii) playing the role of $s$ in (i).
\qed

\begin{thm}\mbox{\ }
Every hereditarily finite set is heretarily closed, and every hereditarily closed set is canonical.  Both implications are strict.
\end{thm}
\proof
The first implication $\HF(s)\Imp\HC(s)$ follows directly from Lemma \ref{lem:cl}(ii) and Theorem \ref{thm:finsing}.

For the implication $\HC(s)\Imp s=F(s)$, one approach would be to show that the binary relation on sets $s,t$ defined by $\HC(s)\wedge t=F(s)$ is a bisimulation.  Alternatively, we can observe that on hereditarily closed sets $s$, the coinductive definition
\begin{eqnarray*}
F(s) &\eqdef& \set{F(u)}{u\in\cl(s)}
\end{eqnarray*}
is equivalent to the coinductive definition
\begin{eqnarray*}
F(s) &\eqdef& \set{F(u)}{u\in s},
\end{eqnarray*}
which uniquely defines the identity function, thus $s=F(s)$ on all such sets.

Both implications are strict.  An hereditarily closed set that is not finite is $\set{\bar n}{n\geq 0}\union\Omega$, and a canonical set that is not closed is $\{\set{\bar n}{n\geq 0}\union\Omega\}$.
\qed

\section{Conclusions and Future Work}
\label{sec:outro}

We have illustrated the use of the metric coinduction principle in four areas: infinite streams, Markov chains, Markov decision processes, and non-well-founded sets.  In all these areas, metric coinduction can be used to simplify proofs or derive new insights.

Other areas are likely to be amenable to such techniques.  In particular, iterated function systems seem to be a promising candidate.

\begin{comment}

\section{Future Work}
\label{sec:future}

Many of these results do not appear to be in their most 
general form.  It may be possible to generalize all 
these application areas in the broader scope of iterated function systems.

An \emph{iterated function system} (IFS) is a complete metric space
$(X,d)$ together with a finite number of mappings $w_1,\ldots,w_n$.
If $d(w_i(x),w_i(y)) < d(x,y)$, then the IFS is 
called \emph{hyperbolic}.  In this case,
the map $T(K)= \bigcup_i w_i(K)$, where $K$ is a 
compact subset of $X$ and $w_i(K)= \set{w_i(x)}{x\in K}$, has a
unique fixed point $A\star$, called the \emph{attractor} of the IFS.

It may be possible to show that certain recursively defined functions 
and Markov chains can be converted to iterated function systems whose 
attractors describe the fixed points of the original system. This may allow 
the above applications to be generalized and analyzed as special cases 
of applying coinductive principles to iterated function systems.
\end{comment}

\nocite{Fel,Barnsley93,IsaacsonMadsen76,PeitgenRichter86,Rutten03,DeBakkerDeVink96}%

\section*{Acknowledgements}

Thanks to Lars Backstrom, Radha Jagadeesan, Prakash Panangaden, and Jan Rutten for valuable comments.  This work was supported in part by ONR Grant N00014-01-1-0968 and by NSF grant CCF-0635028.

\bibliographystyle{plain}
%\bibliography{dk,coinduction,681,types}

\end{document}